\documentstyle[12pt,aasms4,psfig]{article}
\lefthead{Wright et al.}
\righthead{ISO--LWS detection of HD toward the Orion Bar}
 
\begin{document}



\title{ISO--LWS detection of the 112 $\mu$m HD $J$=1$\to$0 line toward the 
Orion Bar$^{1}$}
\altaffiltext{1}{\rm Based on observations with ISO, an ESA project with
instruments funded by ESA Member States (especially the PI
countries: France, Germany, The Netherlands and the United Kingdom)
and with participation of ISAS and NASA.}

\author{Christopher M. Wright\altaffilmark{2,6}, 
Ewine F. van Dishoeck\altaffilmark{2}, 
Pierre Cox\altaffilmark{3}, Sunil D. Sidher\altaffilmark{4} and 
Martin F.\ Kessler\altaffilmark{5}}

\altaffiltext{2}{Leiden Observatory, P.O.\ Box 9513, 2300 RA Leiden, 
The Netherlands}
\altaffiltext{3}{Institut d'Astrophysique Spatiale, Universit\'e de Paris, 
91405 Orsay, France}
\altaffiltext{4}{Rutherford Appleton Laboratory, Chilton, Didcot,
Oxon OX11 0QX, UK}
\altaffiltext{5}{ISO Science Operations Centre, Astrophysics Division
of ESA, P.O. Box 50727, E--28080 Villafranca/Madrid, Spain}
\altaffiltext{6}{School of Physics, University College, ADFA, UNSW, Canberra
ACT 2600, Australia}

\vspace{3 cm}

\begin{abstract}

We report the first detection outside of the solar system of the lowest pure
rotational $J$=1$\to$0 transition of the HD molecule at 112 $\mu$m. The
detection was made toward the Orion Bar using the Fabry-P\'erot of the Long
Wavelength Spectrometer (LWS) on board the {\it Infrared Space Observatory}
(ISO). The line appears in emission with an integrated flux of
(0.93$\pm$0.17)$\times$10$^{-19}$ W cm$^{-2}$ in the LWS beam, implying 
a beam--averaged column density in the $v$=0, $J$=1 state of
(1.2$\pm$0.2)$\times$10$^{17}$ cm$^{-2}$. Assuming LTE excitation, 
the total HD column density is (2.9$\pm$0.8)$\times$10$^{17}$ cm$^{-2}$
for temperatures between 85 and 300~K. Combined with the total warm 
H$_{2}$ column density of $\sim (1.5-3.0)\times$10$^{22}$ cm$^{-2}$ derived 
from either the H$_2$ pure rotational lines, C$^{18}$O observations or
dust continuum emission, the implied HD abundance, HD/H$_{2}$, 
ranges from $0.7\times 10^{-5}$ to $2.6\times 10^{-5}$, with a preferred
value of $(2.0\pm 0.6)\times 10^{-5}$. The corresponding deuterium
abundance of [D]/[H]=$(1.0 \pm 0.3) \times 10^{-5}$ is compared with
recent values derived from ultraviolet absorption line observations
of atomic H {\sc i} and D~{\sc i} in interstellar clouds in the solar
neighborhood and in Orion.

\end{abstract}


\section{Introduction}

The deuterium abundance is one of the most sensitive probes of the baryon
density in the early universe (Wilson \& Rood 1994), since it is thought 
that all the deuterium was produced in the Big Bang, with no subsequent 
production via nuclear reactions in stars. Conversely, it is destroyed in 
stellar interiors, so that current measurements provide only a lower limit 
on the primordial deuterium abundance. Thus, an observed [D]/[H] ratio may 
also be used as a measure of galactic chemical evolution. However, previous 
ground and airplane based attempts to measure the [D]/[H] ratio in a 
variety of interstellar sources have met with several problems, such as 
the very low intrinsic strength of the \hbox{D\,{\sc i}} 92 cm line 
(Heiles et al.\ 1993) or chemical fractionation and line saturation 
effects in molecules (Penzias et al.\ 1977). The most successful 
measurements so far have been through satellite ultraviolet absorption 
line observations of the Ly$\alpha$ lines of atomic \hbox{H\,{\sc i}} and 
\hbox{D\,{\sc i}} through diffuse clouds along the line--of--sight toward 
several stars in the solar neighborhood (i.e. within $\sim$ 90 pc). These 
give [D]/[H]=1.6$\times$10$^{-5}$ with a typical uncertainty of 15\%, a 
value so far independent of the line--of--sight (e.g., Dring et al.\ 1997, 
Piskunov et al.\ 1997). 

Hydrogen deuteride, HD, has also been detected in local diffuse molecular 
clouds through ultraviolet absorption line observations (e.g., Spitzer 
et al.\ 1973), but its abundance is low in these tenuous clouds because 
of rapid photodissociation of the molecule. In contrast, virtually all of 
the deuterium is expected to be contained within HD in dense, warm molecular 
clouds. Measurements of the $v$=0--0 $J$=1$\to$0 R(0) 112.072 $\mu$m line can 
potentially provide a direct and accurate determination of the HD abundance, 
and thereby also the deuterium abundance, in such clouds. A previous attempt 
to observe this line was made by Watson et al. (1985) toward the Orion KL 
region, but resulted only in an upper limit of $1\times 10^{-18}$ W cm$^{-2}$.
In this paper we report the first detection of the 112 $\mu$m HD $J$=1$\to$0
line outside of the solar system.

The detection was made toward the Orion Bar, a warm Photon Dominated Region
(PDR), and the line is observed to be in emission. PDRs have the advantage
over other source types, such as embedded young stellar objects, in that they
have a large amount of warm molecular gas at sufficiently high temperatures to
excite the HD $J$=1$\rightarrow$0 112 $\mu$m line. They do not have cold
surrounding material which could absorb and cancel part of the emission.
The main drawback is that HD is rapidly photodissociated at the edge of the
PDR, so that its abundance does not become large until $A_V\approx 2-3$ mag
into the cloud (e.g., Jansen et al.\ 1995a). The physical and chemical
structure of the Orion Bar is, however, well understood from a variety of
molecular line observations (e.g., Hogerheijde et al.\ 1995, Jansen et al.\
1995b, van der Werf et al.\ 1996). It has a particularly high total H$_{2}$
column density due to its edge--on geometry, facilitating the detection of HD.
Our successful observation provides a determination of the HD abundance, as
well as a probe of the chemistry of molecular clouds. When compared with
observations of molecular hydrogen, it also allows a constraint to be placed
on the cosmologically important [D]/[H] ratio, the first such direct
measurement in a dense molecular cloud without the complicating effects of
chemical fractionation.

\section{Observations and data reduction}

The Orion Bar was observed during revolution 823 using the LWS04
Fabry--P\'erot (FP) mode of the Long Wavelength Spectrometer (LWS, Clegg et
al. 1996) on board the {\it Infrared Space Observatory} (ISO, Kessler et al. 
1996). The rest frequency of the HD $J$=1$\to$0  line is 
2,674,986.66$\pm$0.15 MHz (Evenson et al. 1988) corresponding to a vacuum 
wavelength of 112.072506$\pm$0.000007 $\mu$m. The observed coordinates 
were RA = 05$^{h}$ 35$^{m}$ 20.3$^{s}$ and 
DEC = $-$05$^{\circ}$ 25$'$ 20$''$ (J2000), a position which 
closely corresponds to the peak column density of molecular gas (e.g.,
Burton et al.\ 1990; Parmar, Lacy \& Achtermann 1991; Hogerheijde et al.\ 
1995).
 
The observations consist of 120 separate LWS FP scans centered on the
frequency of the HD $J$=1$\to$0 line, with 7 spectral elements on either side
of the line. The data were taken with the LW2 detector in fast scanning mode,
with 4 spectral samples per resolution element and 45.5 s per sample. The
total on--target--time was 3913 s. The FWHM beam size at 122 $\mu$m, the
central wavelength of detector LW2, in the spacecraft Y--Z directions is
$78''\times75''$ with a systematic uncertainty of up to 20\% (Swinyard et al.\
1998). The resolving power is of order 9500 or $\rm \sim 30 \, km \ s^{-1}$
(Clegg, Heske \& Trams 1994), and the wavelength calibration accuracy is good
to a third of a resolution element, or 10 km s$^{-1}$ (Trams et al. 1998). A
full range 43--197~$\rm \mu m$ LWS01 grating spectrum, consisting of 5 scans
with 0.7 s integration time per step and a spectral sampling interval of 2,
was obtained during the same revolution for calibration purposes (fringes,
continuum level).

Initial data reduction was carried out using the ISO--LWS Off Line
Processing (OLP) software, version 7.0, up to the Auto Analysis Result
stage. Further data processing, such as removal of bad data points,
flat--fielding, sigma clipping and co--adding, was performed using
software in the ISO Spectral Analysis Package, and the LWS and Short
Wavelength Spectrometer (SWS) Interactive Analysis (IA) packages. Of
particular note is the dark current subtraction and grating position
correction. The relatively high flux of the Bar results in some
straylight leakage through the misaligned Fabry--P\'erot plates during
the dark current measurements, meaning that they are not true dark
currents. Therefore, the dark current was iteratively modified such that the
resultant continuum flux was equal to that obtained in the LWS01
observation. Using the LWS IA tool FP\_PROC (Swinyard et al.\ 1998, Sidher
et al., in preparation),
the subsequent product was corrected for a grating positioning
problem, which introduces a spurious slope in the spectrum. By
adopting these procedures, our photometric accuracy is estimated to be
30\% or better (Burgdorf et al.\ 1997, Swinyard et al.\ 1998).

\section{Results}

Figure 1 displays the resulting spectrum after co--addition of all FP scans. 
An emission line is clearly visible at $v_{\rm LSR}\approx$ +13 km s$^{-1}$,
close to the expected $v_{\rm LSR}\approx$ +10 km s$^{-1}$ known for the Bar
(e.g., Hogerheijde et al.\ 1995). The feature is unresolved, with an observed
FWHM of $\sim$ 30 km s$^{-1}$, i.e., near the resolving power of the LWS and
implying an intrinsic FWHM less than this, again consistent with previous
millimeter observations which show $\Delta V\approx 2$ km s$^{-1}$. The
observed integrated line flux is (7.3$\pm$1.3)$\times$10$^{-20}$ W
cm$^{-2}$, obtained by fitting a Gaussian to the line profile, and a 
first order polynomial plus sinusoid due to fringing (see below)
to the baseline. The uncertainty is 
statistical and represents the range of values obtained using different 
fitting procedures. When divided by the LW2 beam size, $1.08\times10^{-7}$ 
sr, the inferred surface brightness is $I=(6.76\pm1.20)\times10^{-6}$ 
erg s$^{-1}$ cm$^{-2}$ sr$^{-1}$. As noted in the LWS Data Users Manual 
(Trams et al. 1998), the LWS OLP software corrects the data of detector 
LW2 by a factor of 0.68 for the effective aperture of the instrument, 
assuming that the source is point--like and located at the aperture centre. 
This is not the case for the Orion Bar, which is extended in both its gas 
and dust emission. The Data Users Manual gives a correction factor of 
0.87/0.68 at 100--120 $\mu$m which must be applied to observed fluxes for 
extended sources. The resulting best estimate of the flux is 
$(9.3\pm 1.7)\times 10^{-20}$ W cm$^{-2}$ and of the surface 
brightness is $(8.7\pm1.5)\times 10^{-6}$ erg s$^{-1}$ cm$^{-2}$ sr$^{-1}$. 
This corresponds to an antenna temperature of $\sim$0.22~K in the LWS beam for 
$\Delta V\approx$2 km s$^{-1}$.

The beam averaged column density in the HD $v$=0, $J$=1 state is obtained from
$N_{\rm 0,1}={(4 \pi I_{1 \rightarrow 0})}/({A_{1 \rightarrow 0} h \nu_{1
\rightarrow 0})}$, where $I_{1\rightarrow 0}$ is the observed surface
brightness. Using $A_{1 \to 0}=5.12 \times 10^{-8}$ s$^{-1}$ (Abgrall, Roueff
\& Viala 1982), $N_{\rm 0,1}$=(1.20$\pm$0.21)$\times$10$^{17}$ cm$^{-2}$ is
found. The small $A$--value assures that the optically thin relation is valid.

Although we are confident that the observed feature is real and corresponds 
to the 112 $\mu$m $J$=1$\to$0 line of HD, there are a few noteworthy points 
to be made. First, it is well known that the LWS grating spectral 
responsivity calibration file at 112 $\mu$m has a spurious ``absorption'' 
feature resulting from HD absorption in the calibration source Uranus. 
However, this does not affect the FP spectrum, where the resolving 
power is a factor of $\sim$ 50 greater than the grating. Across the width 
of our FP spectrum the calibration file is virtually flat. Second, 
a similarly sensitive HD 112 $\mu$m observation was made toward the 
Galactic Center, where no corresponding emission feature is seen
(Wright et al., in preparation).

Another possible source of spurious emission is from leakage into the FP 
of adjacent spectral orders. The FP free spectral range at 112 $\mu$m
is 1.12 $\mu$m, and the LWS Observer's Manual (Clegg, Heske \& Trams 1994) 
discusses possible contamination. The grating acts as an order sorter 
for the FP, and its resolving power was specifically set such that 
contamination is avoided. In any case, at submillimeter wavelengths 
the Orion Bar is very poor in lines compared with spectroscopically 
rich sources such as Orion--KL (Hogerheijde et al.\ 1995), and there
are no other line-rich sources in the LWS beam. The only 
strong lines expected in the LWS wavelength range are the high--$J$ 
transitions from $^{12}$CO and $^{13}$CO. There is no such transition, 
nor any other common low--lying (i.e. $\leq$ 300 K) molecular feature, 
known to be coincident with HD $J$=1$\to$0 within 0.01 $\mu$m, or 
to be a possible contaminant from adjacent spectral orders.

The LWS--FP data are also affected by fringing, in a similar way to LWS 
grating data of extended sources. However, the single high frequency fringing
component apparent in the data has an essentially constant period of about
0.0095 cm$^{-1}$ (Sidher et al.\, in preparation), which makes it easier to
correct. The effect appears to occur as a result of interference within the
instrument.

\section{Analysis and Discussion}

In the following, we will discuss the derivation of the column densities of 
HD and H$_2$ and compare the HD and deuterium abundances in the Orion Bar 
with those found in other regions.

\subsection{The HD column density}

Because of the limited sensitivity of the LWS detector SW2, observations of
the 56.23 $\mu$m 0--0 $J$=2$\to$1 R(1) line of HD were not feasible, so that 
no direct information on the excitation of the molecule is available. 
However, the HD excitation can be readily  computed assuming LTE: 
$N({\rm HD})=[{N_{0,1}Q(T)}/{g_{0,1}}] \exp ({E_{0,1}}/{kT})$, where 
$g_{0,1}$ is the statistical weight of the $v$=0, $J$=1 level, and $Q(T)$ 
is the partition function. Hogerheijde et al.\ (1995) find gas kinetic 
temperatures in the Bar of
85$\pm$30 K from observations of a variety of molecules, whereas other
analyses give values up to at least 300~K (Parmar et al.\ 1991, see \S 4.2).
The corresponding total HD column densities range from 
(2.6$\pm$0.5)$\times 10^{17}$ cm$^{-2}$ at 115--200~K to 
(3.1$\pm$0.6)$\times 10^{17}$ cm$^{-2}$ at 85 and 300~K. Including the 
temperature range as an additional uncertainty, our best estimate of the 
total HD column density is (2.9$\pm$0.8)$\times$10$^{17}$ cm$^{-2}$.
The estimated hydrogen densities of the Orion Bar, such as
10$^{4}$ to 10$^{5}$ cm$^{-3}$ for the so--called interclump medium,  or
$\geq$ 10$^{6}$ cm$^{-3}$ for the high density clumps (e.g., 
Hogerheijde et al.\ 1995), are all above the critical density of the $J$=1
level. For the higher HD levels, the departures from LTE are small 
for these conditions.

\subsection{The HD abundance}

To derive the HD abundance with respect to H$_{2}$, the total H$_{2}$
column density, $N({\rm H_{2}})$, is needed in addition to
$N$(HD). The most direct method is to use observations of the pure
rotational lines of H$_2$ itself.  Parmar, Lacy \& Achtermann (1991)
present data of the 17.0348 $\mu$m 0--0 $J$=3$\to$1 S(1) and 12.2786
$\mu$m 0--0 $J$=4$\to$2 S(2) lines of H$_{2}$ in a $10''\times2''$
slit at positions covering a $10''\times16''$ region in the Bar, and
within our $\sim 76''$ aperture. Our observed position is within a few
arcseconds of their position 3, the point of peak emission, while
their positions 1, 2, 4 and 5 fall well within our aperture. The
averaged line fluxes for positions 1--5 (covering $10''\times10''$)
give a mean excitation temperature $T_{\rm ex}$ between levels $J$=3
and 4 of 482~K and a column density of 9.9$\times$10$^{20}$
cm$^{-2}$. However, this is unlikely to represent the total H$_{2}$
column density, since there is a strong temperature gradient
throughout the PDR with a significant amount of cooler gas present, as
indeed is shown in other molecular tracers.

Several methods may be used to determine the total amount of gas. First, the
clumpy PDR models of Burton, Hollenbach \& Tielens (1990, 1992) can be used to
estimate the expected $T_{\rm ex}$ between levels $J$=2 and 3. For the
conditions appropriate to the interclump medium of the Orion Bar, where Burton
et al.\ (1992) state that the bulk of the H$_{2}$ emission originates, $T_{\rm
ex}$=180--260~K. Using the observed column density in the $J$=3 level, we can
then extrapolate to the $J$=2 level to determine its column density and
thereby the total warm H$_{2}$ column density, assuming that the ortho--para
ratio is in LTE (Sternberg \& Neufeld 1999). For $T_{\rm ex}$=180~K,
$N$(H$_2$)=1.5$\times 10^{22}$ cm$^{-2}$ is found, while for $T_{\rm
ex}$=260~K, $N$(H$_2$)=4.4$\times$10$^{21}$ cm$^{-2}$. We note that recent
observations with the ISO--SWS at the same position in the Bar, including the
28.2188 $\mu$m 0--0 $J$=2$\to$0 S(0) line, show that $T_{\rm ex}$ between 
the $J$=2 and
$J$=3 levels is 150--175~K, and that the total warm H$_{2}$ column density is
$\sim$1.5$\times$10$^{22}$ cm$^{-2}$, with a statistical uncertainty of
$\sim$10\% (e.g. Rosenthal et al., in preparation).

Using $N$(H$_2$)=1.5$\times$10$^{22}$ cm$^{-2}$, the HD abundance,
$N$(HD)/$N$(H$_2$) (denoted as HD/H$_2$), is (2.0$\pm$0.6)$\times$10$^{-5}$.
Our estimate of $N$(H$_2$) may be an underestimate since the H$_2$ S(0) line
may still be weighted toward warmer gas than the HD $J$=1$\to$0 line, given 
that their upper level energies are $\sim$ 510 and 128~K above ground,
respectively. This would lead to an overestimate of HD/H$_2$. This may be
counter--balanced by an intrinsic underestimate due to photodissociation of HD
at the PDR edge and the fact that the LWS beam is significantly larger than
the apertures used for the H$_2$ observations ($14''\times27''$ at 12 and 17
$\mu$m and $20''\times27''$ at 28 $\mu$m for ISO--SWS; see also below).

$N$(H$_2$) has also been estimated from observations of
CO and its isotopic forms to be between 3 and
6.5$\times$10$^{22}$ cm$^{-2}$, varying with precise position, line and beam
size (e.g., Graf et al.\ 1990, Hogerheijde et al.\ 1995, van der Werf et al.\
1996). Using the molecular line maps and the C$^{18}$O 2$\to$1 cut across the
Bar by Hogerheijde et al.\ (1995), the beam dilution in the LWS aperture is
estimated to be a factor of $\sim 0.4-0.5$ compared with their peak emission.
This results in a LWS beam--averaged H$_2$ column density of 
$\sim3\times10^{22}$ cm$^{-2}$ if an H$_2$/C$^{18}$O ratio of $5\times 10^6$ 
is used. The corresponding HD/H$_2$ is (1.0$\pm$0.3)$\times$10$^{-5}$.

This value refers to HD with respect to the total amount of warm and cold
H$_2$. However, because the HD $J$=1 level lies at 128~K above ground it is
only excited efficiently in gas with temperatures above $\sim 30$~K. In
addition, HD is photodissociated at the edge of the PDR, so that the region
over which H$_{2}$ and HD are coexistent is less. These effects can be taken
into account by comparison with detailed physical and chemical models of the
Orion Bar, such as developed by Jansen et al.\ (1995a,b). The models show that
HD becomes abundant 2 visual magnitudes deeper into the cloud than H$_2$, and
that the temperature stays above 30~K up to depths of 5--6 mag. These values 
need to be
convolved with the edge--on geometry of the Bar as outlined by Jansen et al.\
(1995b). At the positions included in the LWS beam, the outer, warm layers are
enhanced significantly, and it is estimated that at least 50\% of the total
column density contains HD at sufficiently high temperatures. Thus, the
relevant $N$(H$_2$) in the LWS beam for comparison with HD is $\sim 1.5\times
10^{22}$ cm$^{-2}$, consistent with that determined from the warm H$_{2}$
above. Similar values are obtained from $^{13}$CO 6$\to$5 data from Lis
(private communication) averaged over a 70$''$ beam and from the high-$J$
$^{12}$CO lines detected in our LWS01 spectrum. The corresponding HD abundance
is again (2.0$\pm$0.6)$\times$10$^{-5}$.

Another method for determining the HD abundance is to utilise the dust 
continuum emission as a tracer of molecular hydrogen. Our LWS01 
spectrum gives a dust temperature of 60~K and a 112 $\mu$m optical depth 
of 0.05, obtained from an optically thin fit to the data. Inserting these 
values into Eq.\ (4) of Watson et al.\ (1985), 
HD/H$_2$=(1.5$\pm$0.3)$\times$10$^{-5}$ is obtained. Alternatively, 
the 400 $\mu$m observations of Keene, Hildebrand \& Whitcomb (1982) can 
be used along with Eq.\ (9) in Table 1 of Hildebrand (1983) to find 
$N$(H$_2$)=2.6$\times$10$^{22}$ cm$^{-2}$ and 
HD/H$_2$=(1.1$\pm$0.3)$\times$10$^{-5}$, using a 400 $\mu$m optical depth
of 4.4$\times$10$^{-3}$. These methods assume 
thermal equilibrium between the gas and dust, however, and are uncertain 
by a factor of 2 due to uncertainty in the dust opacity. If the gas and 
dust are  in thermal equilibrium at $\sim$60~K, $N$(HD) increases to 
$(4.6\pm0.8)\times10^{17}$ cm$^{-2}$ and 
HD/H$_2$=$(1.8\pm0.3)\times10^{-5}$.

Although each of the above methods to derive the HD abundance have their own
intrinsic uncertainties, our observations constrain the range of HD/H$_{2}$ in
the Orion Bar to be between 0.7$\times$10$^{-5}$ and 2.6$\times$10$^{-5}$, not
including the uncertainty due to the LWS photometric calibration and beam size.
Our preferred value is $(2.0 \pm 0.6) \times 10^{-5}$, the value obtained
using the H$_2$ pure rotational lines or corrected CO isotopic emission as
tracers of $N$(H$_2$).

\subsubsection{Comparison with other determinations of the HD abundance}

It is interesting to compare our result with those previously obtained 
toward other source types. The HD/H$_{2}$ ratios obtained from ultraviolet 
absorption line observations through diffuse clouds 
are typically $\sim 10^{-6}$ (e.g. Spitzer et al. 1973, Snow 1975, 
Wright \& Morton 1979). Such low values are explained through preferential 
ultraviolet dissociation of HD, since HD does not self--shield. Although 
gas--phase HD formation is more rapid than that of H$_2$ due to the 
H$^+$ + D $\to$ D$^+$ + H and D$^+$ + H$_2$ $\to$ HD + H$^+$ reactions, 
the net effect is still a low HD/H$_2$ ratio. Future observations with the 
{\it Far Ultraviolet Space Explorer} (FUSE) will be able to extend these
ultraviolet observations to thicker, translucent clouds in the solar
neighborhood, where more of the deuterium is in HD. Recently,
Bertoldi et al.\ (1999) measured the 19.4 $\mu$m HD 0--0 $J$=6$\to$5 R(5) 
line in the Orion shock using the ISO--SWS, and found an HD abundance of  
$(9.0 \pm 3.5) \times 10^{-6}$, corrected for non--LTE population.
In partially dissociative shocks, the HD abundance may be reduced relative
to H$_2$ by $\sim$40\% due to the HD + H $\to$ H$_2$ + D reaction.

The HD abundance has also been determined for the giant planets using the SWS
and LWS on board ISO to measure the 37.7 $\mu$m 0--0 $J$=3$\to$2
R(2) and 56.2 $\mu$m $J$=2$\to$1 R(1)
lines, respectively. These yield (preliminary) values of
(2.6--5.8)$\times$10$^{-5}$ for Jupiter and (3.0--8.0)$\times$10$^{-5}$ for
Saturn (Lellouch 1999, and references therein). Our value is at the low end of
these ranges, perhaps implying some Galactic evolution of the HD (and thereby
D) abundance over the 5 billion year lifetime of our solar system.

\subsection{The deuterium abundance}

Our derived HD/H$_2$ ratios refer specifically to the regions of the
PDR where deuterium is in molecular form. Thus, the amount of atomic
deuterium can be neglected and [D]/[H]$\approx 0.5 \times$
HD/H$_{2}$. Our derived deuterium abundance ranges from 0.35 to 1.30
$\times$10$^{-5}$, with a preferred value of $(1.0 \pm 0.3) \times
10^{-5}$. This result is marginally outside the error estimate
($\sim15\%$) of the value for the solar neighborhood of
1.6$\times$10$^{-5}$ (Piskunov et al.\ 1997). However, it is close to
the value of $(0.76\pm0.29) \times 10^{-5}$ found for the Orion shock
by Bertoldi et al.\ (1999), and $(0.74^{+0.19}_{-0.13})\times
10^{-5}$, $(0.65\pm0.3) \times10^{-5}$ and
$(1.4^{+0.5}_{-1.0})\times10^{-5}$ for the lines--of--sight to the stars
$\delta$, $\epsilon$ and $\iota$ Orionis by Jenkins et al.\ (1999) and
Laurent et al.\ (1979), respectively. Additional ultraviolet and
far-infrared observations are needed to determine whether true
variations in the deuterium abundance exist within Orion and between
Orion and the local interstellar medium. Our value is however
definitely below that of (3.9$\pm$1.0)$\times$10$^{-5}$ found through
radio observations of the 21 cm \hbox{H\,{\sc i}} and 92 cm
\hbox{D\,{\sc i}} hyperfine transitions in a region of reduced star
formation activity in the outer Galaxy by Chengalur, Braun \& Burton
(1997). This provides some evidence of a Galactic deuterium abundance
gradient, and is consistent with its destruction in stars. Better
determinations of [D]/[H] in the Galactic Center will be important to
further constrain the gradient across the Galaxy.

The successful detection of the HD 112 $\mu$m $J$=1$\to$0 line by ISO in a
reasonable integration time implies that the next generation of infrared
airborne observatories, such as the {\it Stratospheric Observatory for
Far--Infrared Astronomy} (SOFIA) and the {\it Far--InfraRed and Submillimeter
Space Telescope} (FIRST), will be able to detect this line in a variety of
sources, and possibly determine a [D]/[H] gradient across the Galaxy. At the
high spectral resolution of the SOFIA and FIRST instruments, other sources
such as dense molecular cloud cores may be more favorable for HD searches,
because of their higher column densities compared with PDRs. The largest
uncertainty in the [D]/[H] analysis remains the determination of the
corresponding H$_2$ column density.

\acknowledgments
 
This work was supported by NFRA/NWO grant 781-76-015. We
thank Frank Bertoldi, John Black, Francois Boulanger, David Jansen, Ed 
Jenkins, Darek Lis, Tom Phillips, Dirk Rosenthal, Ralf Timmermann, Laurent 
Verstraete and Jonas Zmuidzinas for their help and stimulating discussions, 
and the referee John Lacy for instructive comments. 
CMW acknowledges support of an ARC Research Fellowship. 
The ISO Spectral
Analysis Package (ISAP) is a joint development by the LWS and SWS Instrument
Teams and Data Centers. Contributing institutes are CESR, IAS, IPAC, MPE, 
RAL and SRON.

\clearpage

\figcaption[wright_fig1.ps]
{The ISO--LWS Fabry--P\'erot spectrum toward the Orion Bar of the HD 0--0
$J$=1$\to$0 112.0725 $\mu$m line. The presented spectrum has not been
corrected for the effective aperture of the instrument. The error bar
represents the typical standard deviation in individual data points. The
baseline has been corrected for fringing.}


\newpage
\begin{figure}
\begin{center}
\leavevmode
\psfig{figure=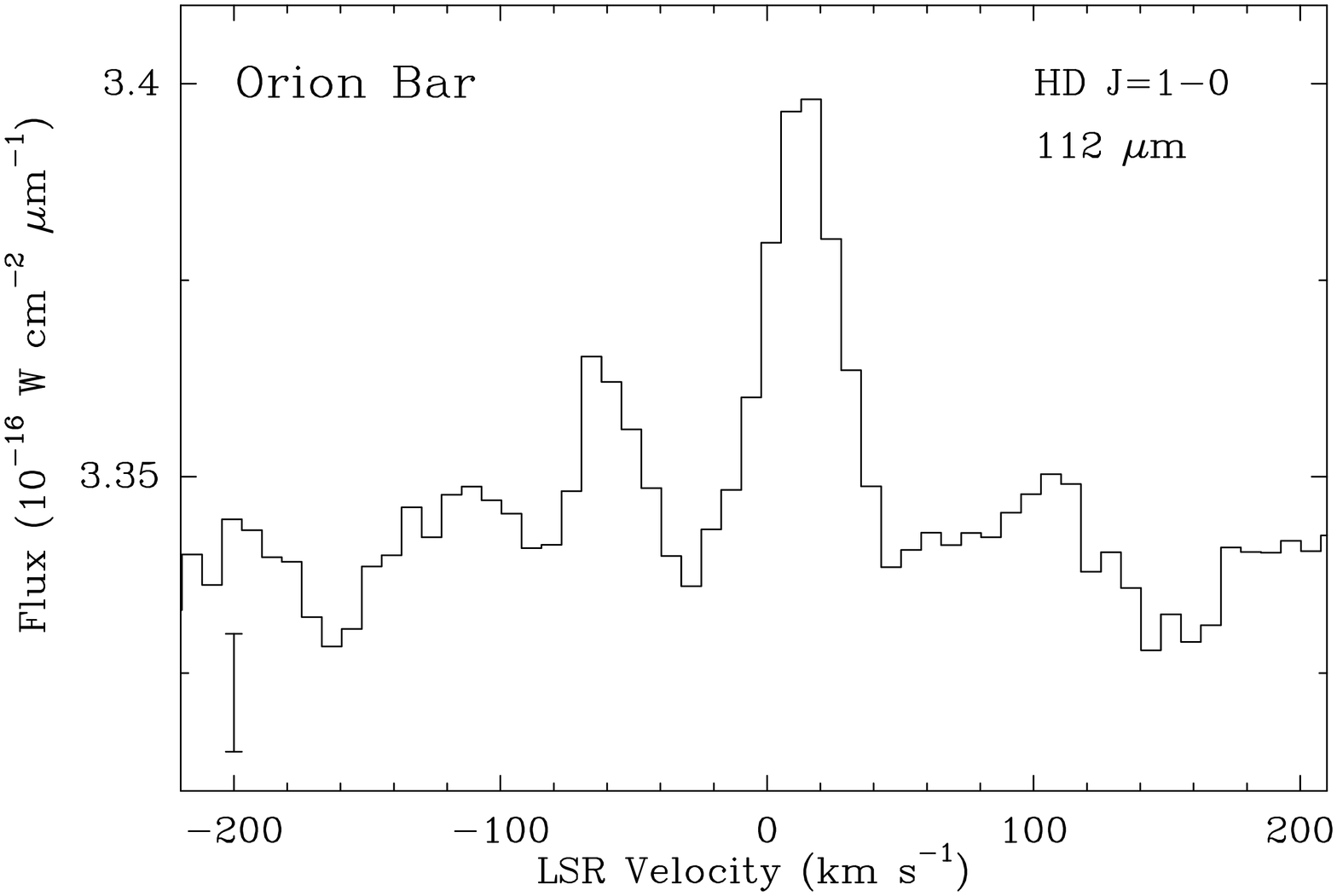,height=10cm,angle=-90}
\end{center}
\end{figure}

\clearpage

\begin{references}

\reference{}
Abgrall, H., Roueff, E. \& Viala, Y.\ 1982, A\&A Supp. Ser., 50, 505

\reference{}
Bertoldi, F., Timmermann, R., Rosenthal, D., Drapatz, S. \& Wright. C.M.\ 
1999, A\&A, submitted

\reference{}
Burgdorf, M., Clegg, P., Ewart, D.\ 1997, in {First ISO Workshop on 
Analytical Spectroscopy}, ESA-SP 419 (Noordwijk: ESTEC), eds. A.M. Heras 
et al., p.\ 51.

\reference{}
Burton, M.G., Geballe, T.R., Brand, P.W.J.L. \& Moorhouse, A.\ 1990, ApJ, 
352, 625

\reference{}
Burton, M.G., Hollenbach, D.J. \& Tielens, A.G.G.M.\ 1990, ApJ, 365, 620

\reference{}
Burton, M.G., Hollenbach, D.J. \& Tielens, A.G.G.M.\ 1992, ApJ, 399, 563

\reference{}
Chengalur, J.N., Braun, R. \& Burton, W.B.\ 1997, A\&A, 318, L35

\reference{}
Clegg, P.E., Heske, A. \& Trams, N.R.\ 1994, LWS Observers Manual,
LWS/PEC/2038.01, Issue 1.0, 6$^{th}$ March 1994, 
{\tt <http://isowww.estec.esa.nl/ISO/iso$_{-}$manuals.html>}

\reference{}
Clegg, P.E., Ade, P.A.R., Armand, C. et al.\ 1996, A\&A, 315, L38

\reference{}
Dring, A.R., Linsky, J., Murthy, J. et al.\ 1997, ApJ, 488, 760 

\reference{}
Evenson, K.M., Jennings, D.A., Brown, J.M., Zink, L.R., Leopold, K.R., 
Vanek, M.D. \& Nolt, I.G.\ 1988, ApJ, 330, L135

\reference{}
Graf, U.U., Genzel, R., Harris, A.I. et al.\ 1990, ApJ, 358, L49


\reference{}
Heiles, C., McCullough, P.R. \& Glassgold, A.E.\ 1993, ApJS, 89, 271

\reference{}
Hildebrand, R.H.\ 1983, Q. Jl. R. astr. Soc., 24, 267

\reference{}
Hogerheijde, M.R., Jansen, D.J. \& van Dishoeck, E.F.\ 1995, A\&A, 294, 792

\reference{}
Jansen, D.J., Spaans, M., Hogerheijde, M.R. \& van Dishoeck, E.F.\ 1995b, 
A\&A, 303, 541

\reference{}
Jansen, D.J., van Dishoeck, E.F., Black, J.H., Spaans, M., \& Sosin, C.\
1995a, A\&A, 302, 223

\reference{}
Jenkins, E.B., Tripp, T.M., Wozniak, P.R., Sofia, U.J. \& Sonneborn, G.\ 1999,
ApJ, submitted

\reference{}
Keene, J., Hildebrand, R.H. \& Whitcomb, S.E.\ 1982, ApJ, 252 L11

\reference{}
Kessler, M.F., Steinz, J.A., Anderegg, M.E. et al.\ 1996, A\&A, 315 L27 

\reference{}
Laurent, C., Vidal--Madjar, A. \& York, D.\ 1979, ApJ, 229, 923


\reference{}
Lellouch, E.\ 1999, in The Universe as seen by ISO, ESA-SP (Noordwijk: ESTEC),
eds.\ P.\ Cox et al., in press


\reference{}
Parmar, P., Lacy, J.H. \& Achtermann, J.M.\ 1991, ApJ, 372, L25

\reference{}
Penzias, A.A., Wannier, P.G., Wilson, R.W. \& Linke, R.A.\ 1977,
ApJ, 211, 108

\reference{}
Piskunov, N., Wood, B.E., Linsky, J.L., Dempsey, R.C. \& Ayres, T.R.\ 1997,
ApJ, 474, 315

\reference{}
Snow, T.P.\ 1975, ApJ, 201, L21

\reference{}
Spitzer, L., Drake, J.F., Jenkins, E.B. et al.\ 1973, ApJ, 181, L116

\reference{}
Sternberg, A. \& Neufeld, D.A.\ 1999, ApJ, in press

\reference{}
Swinyard, B.M., Burgdorf, M.J., Clegg, P.E., Davis, G.R., 
Griffin, M.J., Gry, C., Leeks, S.J., Lim, T.L., Pezzuto, S.,
\& Tommassi, E.\ 1998, in Infrared Astronomical Instrumentation,
SPIE 3354, in press

\reference{}
Trams, N. et al.\ 1998, ISO LWS Data Users Manual, SAI/95--219/Dc, Issue 5.0,
June 1 1998 {\tt <http://isowww.estec.esa.nl/ISO/iso$_{-}$manuals.html>}

\reference{}
van der Werf, P.J., Stutzki, A., Sternberg, A. \& Krabbe, A.\ 1996, A\&A,
313, 633

\reference{}
Watson, D.M., Genzel, R., Townes, C.H. \& Storey, J.W.V.\ 1985, ApJ, 298, 316

\reference{}
Wilson, T.L. \& Rood, R.T.\ 1994, ARAA, 32, 191

\end{references}
\end{document}